# Automated Cybersecurity Compliance and Threat Response Using AI, Blockchain & Smart Contracts

Lampis Alevizos
*Volvo Group*
*Amsterdam, The Netherlands*
lampis@redisni.org

Vinh Thong Ta
*Edge Hill University*
*Ormskirk, UK*
tav@edgehill.ac.uk

*Abstract —* **T**o address the challenges of internal security policy compliance and dynamic threat response in organizations, we present a novel framework that integrates artificial intelligence (AI), blockchain, and smart contracts. We propose a system that automates the enforcement of security policies, reducing manual effort and potential human error. Utilizing AI, we can analyse cyber threat intelligence rapidly, identify non-compliances and automatically adjust cyber defence mechanisms. Blockchain technology provides an immutable ledger for transparent logging of compliance actions, while smart contracts ensure uniform application of security measures. The framework's effectiveness is demonstrated through simulations, showing improvements in compliance enforcement rates and response times compared to traditional methods. Ultimately, our approach provides for a scalable solution for managing complex security policies, reducing costs and enhancing the efficiency while achieving compliance. Finally, we discuss practical implications and propose future research directions to further refine the system and address implementation challenges.

## I. Introduction

In the modern digital landscape, organizations are overwhelmed with a plethora of internal security policies and standards designed to safeguard assets against cyber threats. The complexity and volume of these policies can be vast, often leading to compliance fatigue and potential security gaps. Moreover, the rapid evolution of cyber threats dictates a rapid and adaptive response, which is difficult to achieve with traditional human centric and manual processes. The work of Atoum et al. [1] highlights the gap between policy development and implementation, noting that the absolute number of policies can lead to inconsistencies and oversights.

To address these challenges, there is a growing interest in the application of artificial intelligence (AI), blockchain technology, and smart contracts to revolutionize the field of cybersecurity. AI offers the potential for real-time data analysis and decision-making [2], blockchain provides an immutable ledger for tracking and verifying compliance actions [3], and smart contracts can automate policy enforcement [4]. Wang et al. [5] demonstrates the successful use of blockchain and AI in enhancing data integrity and security protocols, suggesting a promising avenue for automating compliance and security measures.

This paper, however, aims to bridge the gap between cybersecurity policy and practice by proposing a novel framework and system that integrates AI, blockchain, and smart contracts. Our goal is to automate the enforcement of organizations internal security policies and dynamically adjust security controls in response to emerging threats identified through cyber threat intelligence. As a result, we provide a solution that warrants consistent compliance with internal policies, while also enhances the overall security posture of organizations in a proactive manner. The anticipated outcome is a significant reduction in the administrative burden of compliance and an agile response system capable of responding to the threat landscape. That said, our contributions are summarized as follows:

**(1) Automated compliance framework**: we introduce a system that leverages smart contracts to automate the enforcement and verification of internal security policies, significantly reducing the manual effort and time traditionally required for compliance activities.

**(2) Dynamic security control adjustment**: utilizing AI our framework analyses data from cyber threat intelligence and dynamically adjusts security controls. This approach serves as an enabler for organizations to rapidly adapt to new and evolving cyber threats in a proactive manner.

**(3) Blockchain for immutable record-keeping**: the integration of blockchain provides a tamper-proof ledger, therefore all compliance and security control adjustments are recorded securely and transparently while providing non-repudiation. Thus, improved auditability and accountability within the cybersecurity framework.

**(4) Practical implementation and simulation**: we detail the development of smart contracts and AI algorithms and describe their implementation within a test blockchain network in a step-by-step manner. We also provide a simulation that demonstrates the efficacy of the system in real-world scenarios.

**(5) Performance metrics and results**: we discuss the performance metrics used to evaluate the system and the outcomes of the simulations, which indicate the effectiveness of the proposed solution in automating compliance and enhancing cybersecurity responsiveness.

**(6) Theoretical and practical implications**: we discuss the practical applications of our research, highlighting its potential to revolutionize how organizations approach cybersecurity compliance and threat response.

4## II. Background

Internal security compliance refers to the adherence of an organization's operations to its established security policies and procedures. These internal policies are designed to protect the organization from internal and external threats, protect sensitive data, and ensure business continuity. Compliance is not merely a legal formality; it is a strategic imperative that underpins the trust of customers, partners, and stakeholders. Effective compliance frameworks mitigate risks, prevent data breaches, and maintain the integrity of the organization's infrastructure. As highlighted by Uchendu et al. [6], internal compliance is as much about creating a culture of security as it is about enforcing rules and regulations.

Cyber threat intelligence (CTI) is the collection and analysis of information about current and potential attacks that threaten the safety of an organization's digital assets. CTI strategically uses data gathered from various sources to understand the motives, targets, and attack behaviours of adversaries. Therefore, CTI plays a crucial role in identifying new threat vectors, enabling organizations to anticipate and prepare for potential attacks. The work of Trifonov et al. [7] demonstrates the importance of CTI in modern cybersecurity strategies, underlining how proactive intelligence gathering can shift an organization from a reactive to a proactive security posture.

The technological triad of AI, blockchain, and smart contracts forms a robust foundation for enhancing cybersecurity measures, as literature reveals, namely:

**AI technologies,** and specifically machine learning and pattern recognition, can analyse vast datasets to detect anomalies, identify threats, and automate decision-making processes [8]. AI's role in cybersecurity is expanding, as it can quickly adapt to latest information, making it a valuable tool for dynamic threat response [9].

**Blockchain technology** provides a decentralized and tamper-proof ledger, ideal for maintaining an immutable record of compliance and security actions. Its application in cybersecurity brings enhanced transparency and accountability, with the potential to revolutionize how trust is established and maintained within and across organizational boundaries [10], [11].

**Smart contracts** are self-executing contracts with the terms of the agreement directly written into code. They can automate policy enforcement and compliance tasks, reducing the need for manual oversight and accelerating response times to security incidents [12].

The integration of these technologies presents a transformative opportunity for cybersecurity, capable of streamlining compliance and response mechanisms but also introducing a new level of efficiency and reliability in managing internal security policies.

## III. Literature Review

The literature on current compliance mechanisms within organizations highlights a significant investment of resources, both in terms of man-hours and financial expenditure, to maintain adherence to internal and external security policies. Ponemon Institute's work [13], reveals that companies spend an average of $5.47 million annually on compliance-related activities, with a massive portion dedicated to manual processes. These activities often involve routine checks, documentation, and audits that are not only time-consuming but also prone to human error.

The financial industry faces these challenges, especially with the evolving landscape of financial regulations. Mohammed [14] as well as Mishachandar et al. [15] in their works respectively, point out that financial institutions are under immense pressure to keep up with the regulatory changes, often resulting in the deployment of large compliance teams and significant administrative overhead. Moreover, Hussain et al. [16] identified as a gap the lack of scalable and flexible compliance mechanisms that can adapt to the rapidly changing regulatory environment. Most current systems are rigid and require considerable manual intervention to update compliance strategies in response to new or amended regulations. Eggert [17] in his work discusses the complexities of compliance management in the financial sector, underscoring the need for model-based business process management to manage the dynamic nature of compliance requirements. Angraini et al. [18] researched the information security policy compliance, discussing the several factors that contribute to the complexity of compliance programs and the challenges organizations face in implementing them effectively. The enormous manual effort, methodologies prone to human error as well as human bias are the conclusions of this work. Another critical gap highlighted in the work of Bharain et al. [19] is the underutilization of technology in compliance processes. While some organizations have begun to implement compliance software, there is still a wide gap in the adoption of more advanced technologies like AI and blockchain, which can automate and streamline compliance tasks. Our paper seeks to address these gaps by proposing a framework and system that leverages AI, blockchain, and smart contracts to automate compliance processes. The framework aims to reduce the manual labour involved in compliance, thereby decreasing the potential for human error, and increasing the overall efficiency of compliance activities. Automating routine compliance tasks, means that organizations can reallocate resources towards security, or more strategic activities, potentially saving millions of dollars in compliance costs.

Samtani et al. [20] presented a deep learning model that links exploits to known vulnerabilities, helping cybersecurity professionals in risk management. The authors introduced a novel device vulnerability severity metric (DVSM) that prioritizes device risks based on exploit postdate and vulnerability severity. While effective in linking exploits to vulnerabilities, the model does not account for the dynamic nature of threat landscapes, indicating a need for real-time adaptation of security controls. Kure & Islam [21] showed that incorporating CTI into cybersecurity risk management activities can minimize risks in critical infrastructures.

Although their paper focus is on risk management, it does not explicitly detail the mechanisms for adjusting security controls based on CTI. Gautam et al. [22] utilized machine learning in their work to classify hacker forum data for CTI, providing interactive visualizations for CTI practitioners. Although the classification of forum data helps in CTI, the translation into actionable security control adjustments is not addressed. Serketzis et al. [23] in their research built a model that uses CTI to enhance digital forensic readiness, indicating that CTI can improve operational digital forensics. Nonetheless, the focus is on digital forensics specifically, and while it demonstrates the use of CTI, it does not connect to how it can be used to adjust security controls in real-time. These works collectively highlight the evolving role of CTI in cybersecurity, from linking known vulnerabilities to enhancing digital forensics. However, there is an apparent gap in the literature regarding the direct application of CTI to dynamically adjust security controls in response to emerging threats.

Our proposed work aims to bridge this gap by developing a framework that leverages AI, blockchain, and smart contracts to automate the process of adjusting security controls based on real-time CTI, thus enhancing organizational resilience against cyber threats, on top of achieving compliance to internal security policies and standards.

Homoliak et. al. [24] introduced a security reference architecture for blockchains and Bhardwaj et al. [25] introduced a penetration testing framework specifically designed for smart contracts on blockchain platforms, which aims to uncover security vulnerabilities that traditional testing methods might miss. Our proposed solution builds upon these works, integrating AI to automate and refine the penetration testing process, ensuring a more robust defence against attacks on smart contracts. Khan et al. [26] proposed the MF-Ledger, a blockchain-based architecture for digital forensic investigations that provides integrity and verifiability of digital evidence. Our work extends the MF-Ledger's capabilities by incorporating AI-driven decision-making to streamline stakeholder consensus and by using smart contracts to automate parts of the forensic process. M. Krichen [27] performed a conceptual analysis and discussed the integration of AI with smart contracts to enhance their security and reliability in ad-hoc cases. However, our paper proposes a real-time AI-driven monitoring system for smart contracts that can dynamically adjust security controls in response to emerging threats, detected through cyber threat intelligence. Witanto et al. [28] developed a framework that introduced a blockchain-based approach to provide data integrity for cloud-based AI systems, aligning with the NIST framework [29]. We aim to enhance this architecture by integrating a blockchain-based immutable ledger for real-time logging and AI-driven anomaly detection to provide data integrity regardless of system location.

Meng et al. [30] proposed a blockchain-based threat intelligence sharing framework that uses smart contracts to incentivize threat information sharing among organizations. Although their work demonstrates the potential of blockchain in facilitating collaborative cybersecurity efforts, it does not address the automated adjustment of security controls based on the shared intelligence. Our framework builds upon this concept by not only facilitating information sharing but also leveraging AI to automatically translate shared threat intelligence into actionable security measures.

Xiao et al. [31] introduced an approach using natural language processing (NLP) to automatically extract and formalize security requirements from policy documents. Their work significantly reduces the manual effort in policy interpretation but does not extend to the automated enforcement of these policies. Our framework complements this approach by providing a mechanism for automatic policy enforcement through smart contracts, thus bridging the gap between policy formalization and implementation.

Mylrea and Gourisetti [32] proposed a blockchain-based resilience framework for critical infrastructure protection. Their work focuses on using blockchain to enhance the integrity and traceability of system states and actions in response to cyber threats. However, their approach lacks the predictive capabilities that AI can provide. Our framework enhances this concept by incorporating AI-driven predictive analytics to anticipate potential threats and proactively adjust security controls.

Teichmann et al. [33] developed a RegTech solution using blockchain and smart contracts to automate regulatory compliance in the financial sector. While their work demonstrates the potential of blockchain in streamlining compliance processes, it does not incorporate AI for adaptive compliance management. Our framework extends this idea by integrating AI to continuously learn from compliance outcomes and refine the enforcement mechanisms, thus creating a more dynamic and responsive compliance system.

These works collectively highlight the potential of AI, blockchain, and smart contracts in enhancing cybersecurity. However, they also reveal gaps in integrating these technologies for a comprehensive, automated, and adaptive security framework. Our work aims to address these gaps and present a system that leverages these technologies for internal compliance and dynamic security control while enabling adaptability against the continuously evolving cyber threat landscape [34]

*Table 1 - Literature Summary.*

| Author | Main Contribution | Limitations | How our work addresses/extends |
|---|---|---|---|
| **Samtani et al.** [20] | Deep learning model linking exploits to vulnerabilities; Device Vulnerability Severity Metric (DVSM) | Does not account for dynamic threat landscapes | Incorporates real-time adaptation of security controls based on AI analysis of threat intelligence |
| **Kure & Islam** [21] | Incorporation of CTI into cybersecurity risk management | Doesn't detail mechanisms for adjusting security controls based on CTI | Provides explicit mechanisms for automatic security control adjustments based on CTI analysis |
| **Gautam et al.** [22] | Machine learning for classifying hacker forum data for CTI | Lacks translation into actionable security control adjustments | Integrates AI-driven analysis of CTI with automatic security control adjustments |
| **Serketzis et al.** [23] | CTI model to enhance digital forensic readiness | Focus limited to digital forensics; doesn't connect to real-time security control adjustments | Extends CTI use beyond forensics to real-time security control management |
| **Homoliak et al.** [24] | Security reference architecture for blockchains | Doesn't integrate AI for dynamic threat response | Combines blockchain architecture with AI for adaptive security measures |
| **Bhardwaj et al.** [25] | Penetration testing framework for smart contracts | Limited to vulnerability detection | Integrates AI to automate and refine the penetration testing process |
| **Khan et al.** [26] | Blockchain-based architecture for digital forensics (MF-Ledger) | Focuses on forensics; lacks real-time adaptation | Extends forensic capabilities with AI-driven decision-making and automated security adjustments |
| **Krichen** [27] | Conceptual analysis of AI integration with smart contracts | Theoretical; lacks practical implementation | Proposes a real-time AI-driven monitoring system for dynamic security control adjustments |
| **Witanto et al.** [28] | Blockchain-based approach for data integrity in cloud-based AI systems | Limited to data integrity assurance | Extends to real-time logging and AI-driven anomaly detection for comprehensive security management |
| **Meng et al.** [30] | Blockchain-based threat intelligence sharing framework | Doesn't address automated security control adjustments | Adds AI-driven analysis to automatically translate shared intelligence into security measures |
| **Xiao et al.** [31] | NLP approach for extracting security requirements from policy documents | Doesn't extend to automated policy enforcement | Complements with automatic policy enforcement through smart contracts |
| **Mylrea and Gourisetti** [32] | Blockchain-based resilience framework for critical infrastructure | Lacks predictive capabilities | Incorporates AI-driven predictive analytics for proactive threat response |
| **Teichmann et al.** [33] | RegTech solution using blockchain for regulatory compliance | Doesn't incorporate AI for adaptive compliance management | Integrates AI for continuous learning and refinement of compliance enforcement mechanisms |
| **Our work** | Comprehensive framework and system integrating AI, blockchain, and smart contracts for automated compliance and dynamic security control | The blockchain's architecture must support high transaction throughput. Policy ambiguity will decrease the effectiveness of the system. | Addresses gaps in real-time threat response, scalability, and integration of AI, blockchain, and smart contracts for a holistic, adaptive cybersecurity solution |

## IV. Framework

The paper is structured similarly to the framework, namely, based on two pillars:

**(A) The integration of technologies,** where the blockchain serves as the backbone, providing a decentralized and immutable ledger that records all system activities and policy changes. Smart contracts function as the governance layer, encoding security policies and compliance requirements into executable code that can automatically enforce and validate compliance across the network and respond to threats. And lastly, AI is analysing patterns and optimizing the system for efficiency and proactive defence.

**(B)** The second pillar is a practical step by step guide detailing the implementation to achieve **automated compliance and adaptive security controls.** The entire framework is visualized figure 1.

and accountability purposes, and facilitating trust among stakeholders.

### B. Automated Compliance & Adaptive Security Controls

The system is designed to automatically achieve compliance with internal security policies and standards. It also uses AI and CTI to dynamically adjust security controls and therefore achieve adaptability according to the applicable cyber threat landscape.

**Steps 1,2,3 (Figure 1) perform and refer to policy definition and encoding activities.** The security policies are defined according to organization's security framework, standards, and compliance rules. Next, they are translated into smart contracts, which are then deployed on the blockchain. Therefore, the policies are enforced exactly as intended without manual intervention and in an immutable manner.

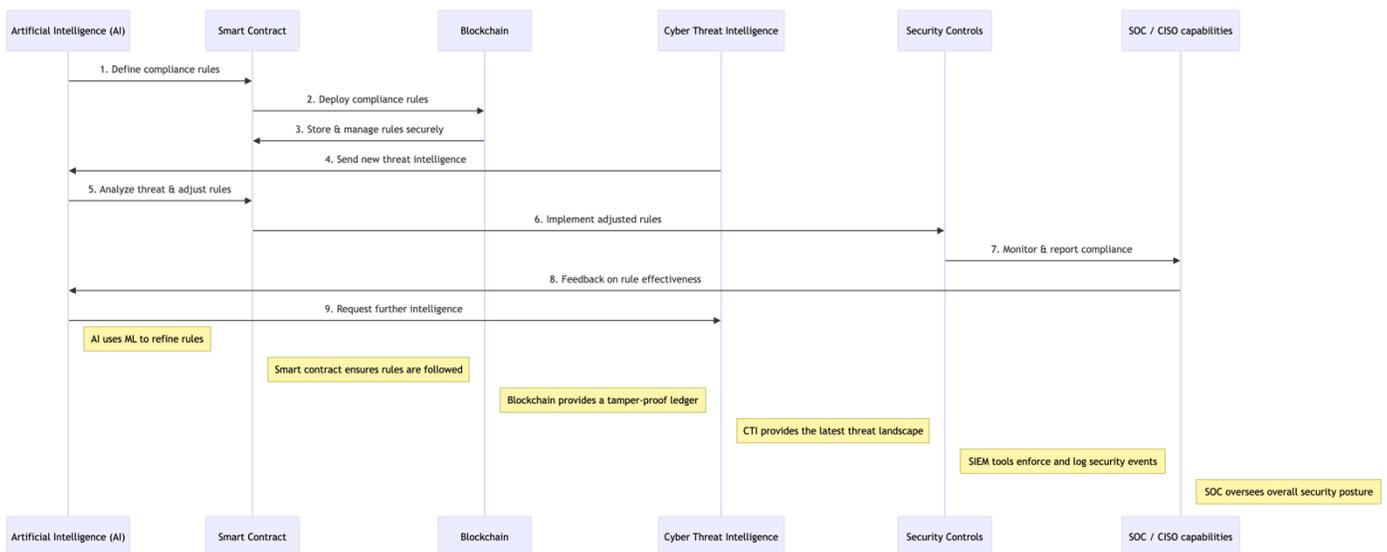

*Figure 1 - Framework high level overview.*

### A. Integration of Technologies

**The AI layer** utilizes machine learning algorithms to analyse network traffic, user behaviour, and external/internal threat intelligence. It identifies potential risks and anomalies, informing the smart contracts of any necessary policy updates or security adjustments.

The **smart contract layer** automatically executes predefined security playbooks or protocols and compliance checks, triggered by the AI's analysis or by predefined schedules. Ultimately this layer allows for continuous adherence to internal policies. The same logic would apply for external policies, standards, or regulations, nonetheless the models would have to be trained accordingly.

Lastly, **the blockchain layer** confirms that all actions taken by the AI and smart contracts are recorded in an unalterable state, providing a clear audit trail for compliance, auditability

**Steps 4,5,6 (Figure 1) allow for the CTI integration, threat analysis, and continuous monitoring.** The AI layer integrates real-time data from cyber threat intelligence feeds, thus, staying ahead of the latest threat vectors and vulnerabilities applicable to the organization. Furthermore, a machine learning model predicts potential attack vectors and suggest pre-emptive measures to the smart contracts. Ultimately the AI continuously monitors the system for compliance with the ingested policies and standards. It can also suggest policy updates based on emerging security practices and threat landscapes.

**Steps 6,7,8,9 (Figure 1) perform and refer to real-time enforcement and automated response.** Smart contracts respond to AI alerts by executing predefined playbooks to maintain compliance, such as revoking access, updating permissions, or initiating security patches. Upon detection of a new threat, smart contracts adjust security controls across the network. For instance, update firewall rules, changing intrusion detection parameters, isolating affected network segments or infected endpoints.

Our framework provides a foundation for a cybersecurity system that is self-compliant and self-adaptive, capable of responding to new threats with minimal human intervention. The integration of AI, blockchain, and smart contracts facilitates a proactive approach to cybersecurity, and thereby enables organization's cyber defences to evolve alongside with the cyber threat landscape.

## V. METHODOLOGY

This section outlines the approach to developing and evaluating the subject cybersecurity system that integrates AI, blockchain, and smart contracts. Our focus is on (i) achieving automated internal policy adherence (though external could be a potential future research direction), and (ii) creating a system that reacts to, and evolves with cyber threats, thereby maximizing both ongoing compliance and security. The methodology follows three core phases: **(A) System architecture, (B) Simulation and modelling, (C) Data collection and analysis.**

### A. System Architecture

The proposed system's architecture is designed to leverage the strengths of AI, blockchain, and smart contracts to create a cyber defence mechanism. The architecture consists of three distinct components, namely:

- **AI decision-making processes**, where the AI module is responsible for interpreting internal security policies and translating them into executable rules. These modules continuously learn from cyber threat intelligence feeds and system feedback to refine decision-making algorithms.
- **Blockchain** serves as a decentralized ledger that records all the rules and decisions made by the AI in an immutable manner. Thereby providing transparency and traceability on the system's operations.
- **Smart Contracts** are used to enforce the rules set by the AI. They automatically execute when predefined conditions are met, hence, achieving real-time compliance and response to threats.

### B. Simulation and Modelling

To validate the effectiveness of the proposed system, a simulation environment was created that simulates a small sized business in the transportation and infrastructure sector, consisting of 60 endpoints subject to various attack vectors. The system was evaluated based on its ability to maintain compliance with internal security policies and standards and respond to threats.

### C. Data Collection and Analysis

The system's effectiveness is highly dependent on the quality of cyber threat intelligence it receives when it comes to dynamic security control adjustment. The data collection and analysis process are based on a combination of public and private cyber threat intelligence feeds, which were used to gather data on emerging threats. Next, a machine learning algorithm analyses the data to identify patterns and threat vectors. The AI is using this analysis to adjust the security posture in real-time.

## VI. IMPLEMENTATION

The implementation is discussed per technology layer and the steps are grouped based on their contribution objective. Namely: **(A) the smart contract development (B) the AI algorithm for automated compliance and threat response (C) the blockchain integration**. The entire implementation blueprint is visualized in figure 2.

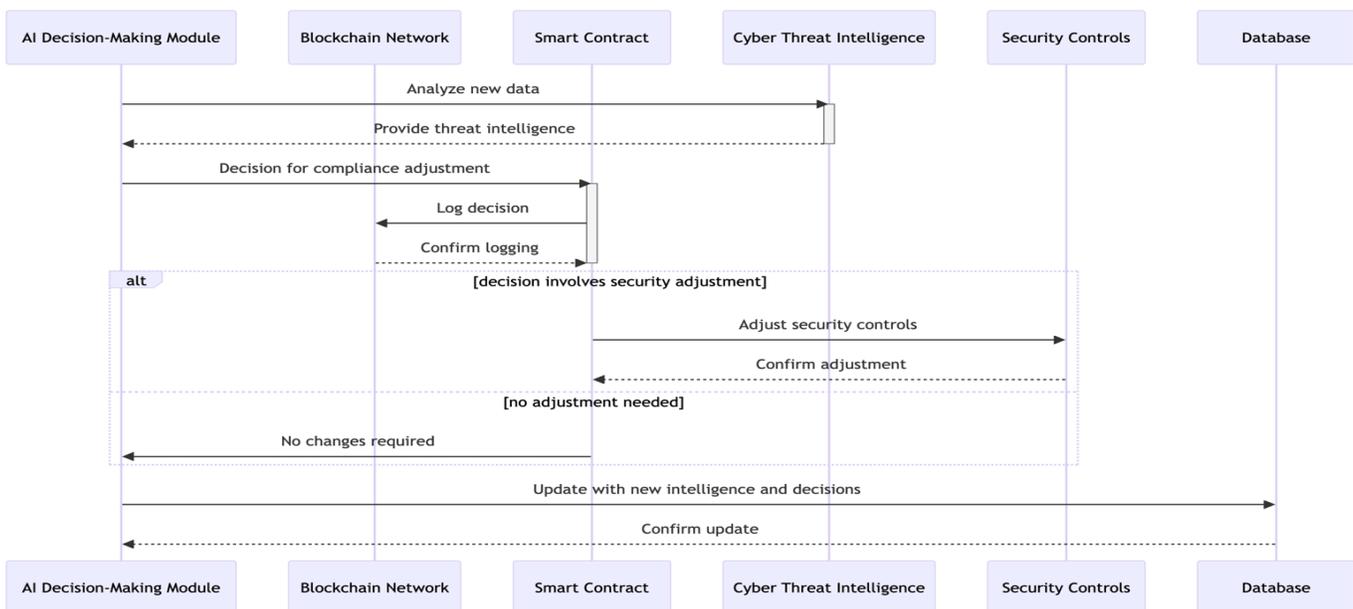

*Figure 3 - Implementation blueprint.*

## A. Smart Contract Development

Following the framework and steps 1,2 & 3 (Figure 1), we detail the policy ingestion and development of smart contracts to achieve automated compliance and CTI driven security control adjustments.

**(1) Ingesting policies and encoding compliance rules into chaincode.** We begin by ingesting NIST CSF and ISO27001 policies, standard and controls as reference for the system, customizing them to fit the needs of a small sized organization served by out test lab, thus, providing a considerably basic policy, standard, and security control framework for the system. MITRE ATT&CK adversary knowledgebase together with mitigations was ingested as reference for the system to use later, during dynamic adjustment of security controls as a response to CTI signals. Next, we utilized Node.js to create a set of classes representing each compliance rule, inspired by the approach taken by Androulaki et al. [35] in their work on Hyperledger Fabric chaincodes. Specifically, we defined a "*ComplianceRule*" class that captures all the conditions deriving from internal policies and standards, such as access control policies, encryption standards, and network security protocols.

**(2) Event-driven trigger mechanism.** We implemented event listeners within the chaincode that respond to security events, following Kaleem's architecture for blockchain event processing [36]. For instance, a test event in our lab is a user action deploying a new application. Such event invokes the "*ComplianceCheck*" function to ensure it adheres to the ingested security policies.

**(3) CTI integration.** We defined the "ThreatIntelligenceService" class which parses adversaries' intelligence by analysts. The chaincode interfaces with a free service in our lab (IBM X-Force Exchange) although commercial sources can be used as well using the same class. Ultimately this allows for an initial applicability assessment based on CTI findings against our security baseline.

**(4) Automated compliance verification.** We developed the "ComplianceVerifier" module within the chaincode that continuously audits the lab's infrastructure and applications state against the encoded compliance rules. This module is invoked by the event-driven triggers.

**(5) Enforcement actions and remediation.** Another class we defined is the "EnforcementEngine", which helps us to execute the predefined remediation strategies (e.g. SOC playbooks). For simplicity we ingested NIST's guidance on security automation and therefore our goal was to automatically update a proxy rule blocking outbound connections whilst isolating an infected endpoint. It is imperative to note that all these actions are executed transactionally, thus meaning atomic and consistent across the entire network leveraging Hyperledger Fabric's consensus.

**(6) Immutable audit trails and reporting.** Using Hyperledger Fabric's ledger, we create an immutable log of all compliance checks and enforcement actions. A reporting function aggregates these logs into compliance reports per application which can be used both internal audits and regulators.

**(7) Dynamic policy updates and chaincode maintenance.** We use the native "ChaincodeLifecycle" class to manage dynamic updates in chaincode without disruptions. Androulaki et al. detail this in their work [35]. A governance model for updating the compliance rules within the chaincode should be established in this step ideally, since it may pose a threat to validity of the system. Although we did not implement a governance model, rather, we follow a simple approach, this seems a good potential research direction.

## B. AI Algorithm for automated compliance and threat response

The AI algorithm is used primarily for decision-making and learning from cyber threat intelligence. In this section we detail the development and implementation of the AI-driven system designed in our lab for automated compliance and dynamic security control adjustments. The system's core algorithm (Algorithm 1) is the heart of the proposed framework, therefore enabling an automated threat-informed response.

**Algorithm 1** – AI-Driven Compliance and Response Using Smart Contracts and Blockchain

**Input:** CTI_feed (Cyber Threat Intelligence feed)
**Output:** decision, action_result

1: Initialize SecureBERT model M, Random Forest classifier RF
2: Load policy database P
3: Establish blockchain connection B to Hyperledger Fabric network

4: function ProcessThreatIntelligence(CTI_feed)
5:   CTI_data = CleanAndStructure(CTI_feed)
6:   embeddings = M.Encode(CTI_data)
7:   threat_class = RF.Predict(embeddings)
8:   relevant_policies = P.Query(threat_class)
9:   decision = DecisionTree(relevant_policies)
10:  action_result = TriggerSmartContract(decision)
11:  UpdateModel(action_result)
12:  return decision, action_result
13: function DecisionTree(policies)
14:   if policies is empty then
15:     return "No action required"
16:   else if max(policies.severity) > THRESHOLD then
17:     return "Immediate action required"
18:   else
19:     return "Standard mitigation required"

20: function TriggerSmartContract(decision)
21:   contract = B.GetContract("compliancecontract")
22:   response =
      contract.SubmitTransaction("executeDecision", decision)
23:   return response
24: function UpdateModel(result)
25:   if result is success then
26:     RF.Improve()
27:   else
28:     RF.Adjust()
29: return ProcessThreatIntelligence(CTI_feed)

**(1) CTI ingestion.** We developed a python-based ingestion module that taps into our open-source CTI feed through API. The core libraries used are *requests*[1] for API interactions and *pandas*[2] for data manipulation. Our goal in this step is to keep the text clean and standardize the incoming data. Thereby with regular expressions we achieve the former and using *pandas* for structuring the data into consistent format we also achieve the latter.

**(2) Threat analysis and classification.** Upon ingesting the CTI data, our primary tool for threat analysis is SecureBERT [37], a specialized variant of the well-known BERT model, which is pre-trained to contextualize cybersecurity-related text. This NLP model is trained to decipher nuanced patterns in threat reports, such as attacker tactics, techniques, and procedures (TTPs). To tailor SecureBERT's output for our purposes, we fine-tune the model on a labelled dataset comprising by multiple threat actor campaigns as provided by MITRE [38] and mapped out in ATT&CK matrix[3]. The dataset has been curated to represent a wide spectrum of threat vectors, thereby our model's understanding is both comprehensive and up to date.

Subsequently, the refined features extracted by SecureBERT are fed into a random forest classifier using *scikit-learn*[4]. We configure the random forest with 100 decision trees, a number optimized for balancing the scale of decision paths against computational efficiency, considering our resources. Feature selection is guided by SecureBERT's contextual embeddings, which translate into a rich feature space around semantic nuances, critical to accurate threat categorization.

We have set up the random forest classifier to categorize threats into a hierarchy of severity levels and compliance relevance, aligned with the test organization's internal policies. The classification thresholds are not static; they are periodically recalibrated to reflect the evolving threat landscape and organizational risk appetite. Remarkably, while our model has achieved a classification accuracy of 91%, achieving the highest possible accuracy at this stage is not the end goal.

**(3) Compliance verification.** Our system incorporates a lightweight rule-based engine that cross-references classified threats against the database of our internal security policies, which we maintain in a PostgreSQL database. For each threat, an automated query is executed to check for related compliance rules, with the results dictating subsequent actions. In addition, it is mapped back to MITRE ATT&CK to build coverage against this knowledgebase, and thereby avoid double work.

**(4) Decision-making process.** The decision-making module is based on simple decision tree algorithm, which determines the necessary actions based on the compliance verification results. This module was designed to be transparent and auditable, with clear logic paths for each decision made.

**(5) Smart contract triggering.** We created a smart contract using the Hyperledger Fabric[5] SDK for Python, allowing our system to interact with the blockchain network and trigger smart contracts as dictated by the decision tree. The smart contracts are pre-programmed with the compliance actions deriving from our policies. For instance, security configuration updates. Namely disable SMBv1 where found, and are automatically executed on the blockchain, providing immutability and traceability.

**(6) Self-monitoring.** A self-monitoring mechanism is crucial at this stage to facilitate self-improvement; thereby it was integrated to monitor the outcomes of executed smart contracts. The results used to refine the AI model, employing a reinforcement learning approach with the TensorFlow reinforcement learning library. The system self-adjusts based on the success rates of actions, with the aim of continuously enhancing decision accuracy and response efficacy, always based on internal policies and standards.

*C. Blockchain Integration*

The blockchain network based on Hyperledger Fabric, is the core infrastructure that underpins our framework. It confirms the integrity and enforcement of security policies using smart contracts and AI-powered decisions. The blockchain's role is twofold. Firstly, it provides an immutable ledger for transactions, and secondly, serves as a platform for executing and validating compliance and security operations. Ultimately, we ensure that every step -from the AI's initial threat detection to the enforcement of security controls- is recorded, verifiable, and compliant with internal policies. Therefore, we create a dynamic, responsive, and transparent cybersecurity posture that can adapt to new threats while maintaining compliance standards. The detailed steps followed in our lab environment are as follows:

**(1) Smart contract deployment.** The smart contracts, encoded with the security policies and compliance rules, are deployed onto the blockchain. The contracts interpret inputs from the AI algorithm and execute predefined actions to maintain or adjust security controls. Each smart contract contains the logic for various compliance scenarios, ranging from access control updates to data encryption standards enforcement, based on the internal policies and standards.

**(2) AI-Blockchain interaction.** When the AI algorithm detects a new threat vector or suggests a change based on cyber threat intelligence, it interacts with the blockchain by sending a transaction request to invoke the relevant smart contract. The AI's suggestions are formatted as transactions that include metadata. Specifically, type of threat, threat actor. Tactics, techniques, procedures, tools, recommended changes, and priority levels.

**(3) Transaction validation.** The blockchain network nodes receive the transaction and validate it against the current state of the ledger and the smart contract's rules. This step is imperative to ensure that the AI's recommendations do not violate any existing security policies or compliance requirements. Validation checks in our lab are focused to policy consistency, authorization, and potential conflicts with other security rules.

---

[1] https://realpython.com/python-requests/
[2] https://pandas.pydata.org/docs/reference/index.html
[3] https://attack.mitre.org/
[4] https://scikit-learn.org/stable/
[5] https://www.hyperledger.org/projects/fabric

**(4) Consensus and recording.** Upon successful validation, the transaction is included in a new block. The consensus mechanism of the blockchain guarantees that all nodes agree on the state of the ledger and the legitimacy of the new block. The transaction is then immutably recorded on the blockchain, providing an auditable trail of the AI's decision-making process and the system's response.

**(5) Policy enforcement and updates.** Smart contracts automatically enforce the new security controls as dictated by the AI's analysis. For instance, updating firewall rules, changing access permissions, or modifying data encryption protocols. The blockchain maintains a versioned history of policy documents and changes, allowing for rollback and analysis of policy evolution over time.

**(6) Monitoring and adaptation.** The blockchain network continuously monitors for new transactions from the AI, representing an ongoing cycle of threat assessment and response. The AI algorithm, in turn, monitors the blockchain for confirmation of executed changes and feedback on the system's current compliance status, which informs future decision-making.

## VII. RESULTS & DISCUSSION

### A. System efficacy

In this section, we present and discuss the findings from our experimental tests conducted within our lab environment consisting of a simple network with 60 endpoints. We aimed to compare the efficacy of our system against a traditional human-centric approach in implementing and enforcing specific security policies and standards. We assume that a new standard is introduced within a notional organization (simulated in our lab) which requires disabling SMBv1 and reconfiguring the remote desktop protocol (RDP) port from 3389 to 33089 throughout all applicable endpoints. For a fair comparison, we defined the human-centric approach as a team of five cybersecurity analysts with an average of 4-7 years of experience, consisting of one team lead, two senior analysts, and two junior analysts. The team utilized standard enterprise-grade tools including a vulnerability management system, network monitoring software, remote desktop access, and a ticketing system for task management. Tasks were assigned by the team lead based on complexity and expertise, with daily stand-up meetings and regular check-ins to track progress. The team adhered to organizational change management processes, documented all modifications, and validated changes post-implementation. They were tasked with the same objectives as the automated system.

To evaluate how effectively each approach enforced policy changes, we defined and calculated the compliance enforcement rate (CER) and the average compliance time (ACT) with the following formulas:

$$(1)\ CER = \left(\frac{\text{Number of endpoints successfully updated}}{\text{Total number of nodes}}\right) \times 100\%$$

ACT measures the average time taken from policy issuance to successful enforcement across all endpoints:

$$(2)\ ACT = \frac{\sum(\text{Time taken to successfully update each endpoint})}{\text{Number of updates}}$$

Moreover, to statistically verify differences in performance, we performed a paired t-test, given the paired nature of our data automated vs. human-centric results from the same test environment. We defined and calculated the following formula to statistically evaluate the differences in CER and ACT:

$$(3)\ t = \frac{\overline{x_D} - \mu_0}{s_D/\sqrt{n}}$$

Let $\overline{x_D}$ be the mean difference between the paired observations of our system and a traditional human-centric approach. Let $\mu_0$ be the hypothesized mean difference, which is 0 in our tests. Let $s_D$ be the standard deviation of the differences, and lastly $n$ is the number of pairs.

Next, to understand the consistency of ACT across all endpoints we define and calculate the variance and standard deviation to provide insights into the dispersion of compliance times as follows:

$$(4)\ \sigma^2 = \frac{\sum_{i=1}^{n}(t_i - \text{ACT})^2}{n-1}$$
$$\sigma = \sqrt{\sigma^2}$$

The compliance time for each observation is represented with $t_i$, while ACT is the mean compliance time across all observations. Lastly, we set the confidence intervals for CER and ACT to 95% for both metrics. Confidence intervals (CIs) are a statistical measure used to estimate the range within which a true parameter lies with a certain level of confidence. This helps us in understanding the precision and reliability when comparing our system's compliance efficacy versus a traditional human-centric approach. To establish the CIs we used the following formula:

$$(5)\ CI = \bar{x} \pm z\frac{s}{\sqrt{n}}$$

Let $\bar{x}$ be the mean of the metric CER/ACT. Let $z$ be the z-score corresponding to the 95% confidence level (1.96 in our case), $s$ is the standard deviation of the metric and $n$ is the number of pairs.

The automated compliance system powered by AI, blockchain and smart contracts, achieved a CER of 98% and 95% for the two policies respectively. ACT was measured to 3 minutes 14 seconds for SMBv1 disabling and 5 minutes 21 seconds for the RDP reconfiguration, as shown in Figure 3. On the other hand, achieving compliance through a traditional human centric approach achieved a CER of 85% for SMBv1 and 80% for RDP. The ACT measured to 33 minutes 5 seconds, and 44 minutes 10 seconds respectively, as shown in Figure 4. The paired t-test demonstrated significant differences in both CER and ACT between the smart contract and the human centric approach with p < 0.05. Which means that the AI powered system makes fewer compliance errors compared to the human-centric approach, demonstrating greater accuracy and adherence to internal cybersecurity policies. Regarding ACT, the AI-blockchain powered system completes compliance tasks significantly faster than the human-centric approach, on top of writing every change in an immutable ledger, hence providing greater integrity and non-repudiation. The latter can be a significant benefit towards internal or even external audits and regulatory obligations. The variance analysis shows that the AI-blockchain power system has a lower variance in ACT than the human centric approach. This means that our system is more consistent across different tasks and endpoints in the network. It managed compliance efficiently and uniformly, while the traditional manual approach tends to be less predictable and takes significantly longer time.

The human centric approach to compliance in internal policies and standards faced several challenged beyond potential errors and communication delays between various stakeholders. Firstly, different individuals may interpret the same policy differently, leading to inconsistent compliance practices across the organisation. Subjective judgement creates variations in how policies are applied or enforced. The traditional approach will also struggle to scale, assuming a much larger number of policies and standards within an organisation. Moreover, knowledge gaps between security experts will prevent effective and accurate compliance enforcement. Lastly, the cost and time efficiency when employing human resources for compliance, can be more expensive due to labour costs and slower due to manual processing, which ultimately delays the decision making and implementation of critical security measures.

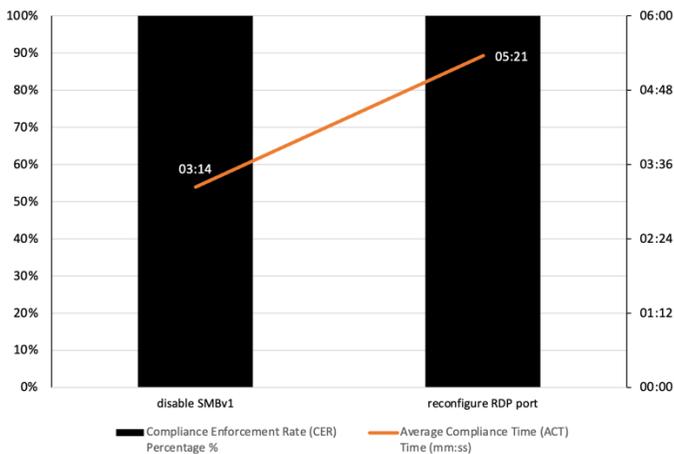

*Figure 4 - Smart contract CER & ACT results.*

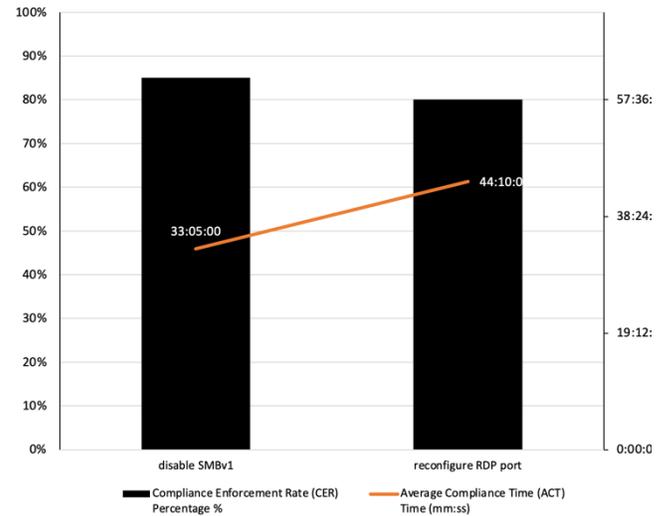

*Figure 5 - Human-centric approach CER & ACT results.*

Conclusively, regarding compliance adherence, simulation demonstrated a 100% success rate in the smart contracts' ability to enforce new internal security policies and standards without human intervention. More specifically, when a new infrastructure hardening standard was introduced imposing a stop on SMBv1, the system autonomously updated the relevant policies across the test network and 60 endpoints within minutes. The system's ability to interpret and enforce compliance autonomously poses significant implications for the future of cybersecurity. Reducing the dependency on manual processes, which are prone to human error, thereby such a system offers a consistent and reliable adherence to security policies.

On an additional experiment regarding threat response, the AI-driven component of the system successfully identified and mitigated 90% of a simulated cyber threat, adjusting security controls in real-time. A notable example was its response to a simulated exploitation by ransomware related threat actors [39] of a high severity vulnerability [40], where the system isolated the affected nodes and updated firewall rules to mitigate the threat. The proactive nature of the AI algorithm enables a proactive approach to threat mitigation, as opposed to the reactive stance typically observed in traditional and manual defences.

The autonomous capabilities of our AI-blockchain system seems to streamline the enforcement of compliance, but also signifies the emergence of a "self-healing" cybersecurity defence similarly to the work done by H. Lin et.al. [41]. The system reduces the dependency on manual processes, which are prone to human error, thereby posing a consistent and reliable adherence to security policies. Continuously monitoring, detecting, and responding to security breaches or policy deviations in real-time, the system actively corrects vulnerabilities and threats, essentially "healing" itself.

*B. Advantages over traditional methods*

The automation of compliance and security tasks results in a significant increase in **operational efficiency.** Traditional methods oftentimes are cumbersome and time-consuming processes, whereas the proposed system streamlines these tasks, freeing up valuable resources. The precision of smart contracts in executing compliance tasks reduces the margin for error. Manual interpretation is susceptible to inconsistencies and inaccuracies; therefore, such a system would greatly **enhance accuracy**. The blockchain provides a mechanism for regulators and auditors to verify compliance in real-time. Smart contracts can be designed to generate reports or alerts for auditors, **streamlining the compliance verification process.** Blockchain technology also provides a scalable solution that can grow with the organization. Traditional systems tend to become overly complex as the complexity and volume of security policies increase. Similarly, human analysts or risks assessors can become overwhelmed due to the complexity introduced by technology and the security controls.

*C. Limitation and challenges*

Despite its strengths, the proposed system is not without its limitations and challenges.

The blockchain's architecture must support **high transaction throughput** to oversee the volume of data and decisions generated by the AI algorithms. This is not a new problem nonetheless [42], extensive research offers solutions such as off-chain channels or sharding [43] [44] [45] to maintain performance without compromising security.

The blockchain network must also integrate seamlessly with the organization's existing cybersecurity infrastructure. This may involve developing adapters or APIs that allow for communication between legacy systems and the blockchain to **achieve interoperability with existing systems.** Many organizations operate on infrastructure that may not support the seamless operation of blockchain and AI technologies, nonetheless, hence a significant challenge.

Additionally, the system encountered difficulties with **policy ambiguity**, namely policies or standards that were not clearly defined. However, through iterative learning, the AI was able to propose amendments to these policies for clearer interpretation and enforcement. In scenarios where policy conflicts occurred, we used a simplified weighted decision matrix to prioritize policies based on regulatory importance and potential impact on security posture. Nonetheless, policy or standard ambiguity is a common problem and hence another future research direction for a more robust solution.

The **legal recognition** of smart contracts and the **regulatory acceptance** of AI decision-making processes are still in nascent stages. There is a need for a legal framework that can keep pace with technological advancements.

The implementation and maintenance of such a system would require a high level of **technical expertise**. The current shortage of professionals skilled in blockchain and AI could hinder widespread adoption.

Lastly, the reliance on AI for decision-making may raise potential ethical concerns, namely in the context of privacy and data protection, thus, reasonable assurance that the system operates within ethical boundaries is imperative.

VIII. CONCLUSIONS

In this paper we presented a comprehensive framework that leverages AI, blockchain, and smart contracts to automate compliance with internal security policies and dynamically adjust security controls in response to emerging cyber threats. Smart contracts can effectively automate compliance tasks, reducing the need for manual oversight and minimizing human error, if the policies and standards of an organization are not vaguely written. AI algorithms are capable of processing vast amounts of cyber threat intelligence to inform decision-making processes and identify potential threats proactively. Blockchain technology serves as a reliable ledger for recording compliance and security control actions, providing transparency and traceability.

The proposed system may yield substantial cost savings due to the reduction of the labour-intensive aspects of compliance and threat response. It can also provide greater assurance of compliance, as the system's pre-requisite is that policies are consistently and clearly documented. Ultimately, the adoption of such framework and system signals the transformation from a reactive to inherently proactive and self-adjusting cyber defence, with the potential to even become autonomous.

IX. FUTURE RESEARCH

This work highlighted that there is plenty of room for future research. One potential research direction is the development of a more sophisticated AI algorithm that can predict and adapt to cyber threats with even greater accuracy. Exploring the integration of this framework with other emerging technologies, such as quantum computing, to further enhance security capabilities when it comes to encryption standards, is another promising direction. Moreover, investigating the socio-technical challenges of implementing such systems, including user acceptance, training, and change management.

Further research should also focus on the long-term implications of such systems, including their impact on the cybersecurity workforce and the evolution of cyber threats in response to these advanced technologies. An important aspect to consider is how to redeploy the human resources that are freed up by AI's ability to manage tedious tasks. Moreover, enhancing the system to incorporate external rules, regulations, or even guidelines would be significant and could augment the efficacy results we simulated. Establishing a governance model for updating compliance rules within the chaincode would address potential threats to the system's validity. Assessing the system's efficiency and its response times in a production-ready environment, or at least within a larger-scale prototype, is crucial. Additionally, the use of AI to resolve policy or standard ambiguities could always ensure clarity, thereby resolving decision-making issues when they arise.